\documentclass[prl,amsmath,amssymb,twocolumn]{revtex4}
\usepackage{graphicx}

\newcommand{\ket}[1]{\mbox{\ensuremath{|#1\rangle}}}

\begin{document}

\title{Experimental quantum key distribution based on a Bell test}

\author{Alexander Ling $^{1,2}$}
\author{Matthew P. Peloso $^1$}
\author{Ivan Marcikic}
\author{Valerio Scarani $^{1,3}$}
\author{Ant\'{\i}a Lamas-Linares $^{1,3}$}
\author{Christian Kurtsiefer $^{1,3}$}

\affiliation{$^1$ Department of Physics, National University of
Singapore, Singapore 117542\\$^2$ Temasek Laboratories, National University
  of Singapore, Singapore 123456\\$^3$ Centre for Quantum Technologies,
  National University of Singapore, Singapore 117543
}

\date{\today}
\begin{abstract}
We report on a complete free-space field implementation of a modified Ekert91
protocol for quantum key distribution 
using entangled photon pairs. For each photon pair we perform a random choice
between key generation and a Bell inequality. The amount of violation 
is used to determine the possible knowledge of an eavesdropper to ensure
security of the distributed final key.
\end{abstract}


\maketitle
{\it Introduction.}
Proposals for quantum key distribution (QKD) were first published over two
decades ago \cite{bb84,quantummoney,bb84expt}. 
In particular, the protocol of Bennett and Brassard in 1984 (BB84) sought to
distribute a random encryption key via correlated polarization states of
single photons \cite{bb84,bb84expt}.  
Its strength was derived from the no-cloning theorem
\cite{nocloning1,nocloning2} which states that the
state of a single quantum system cannot be copied perfectly. 
A measurement attempt on the distributed key is revealed as errors in the
expected correlation of the measurement results. BB84 must treat all noise as evidence of an eavesdropper.
Whether a completely secure key can then be distilled after error correction \cite{bennett95} depends only on the fraction of errors in the initial key.

The `quantum' nature of QKD was explored from a different angle in 1991 when
Ekert proposed an implementation using non-local correlations between
maximally entangled photon-pairs \cite{E91}. 
The quality of entanglement between a photon-pair can be measured by the
degree of violation of a Bell inequality \cite{bell66}. 
Maximally entangled photon-pairs have perfect correlations in their
polarization states, and violate the  Clauser-Horne-Shimony-Holt (CHSH)
version \cite{chsh} of this inequality with the maximum value.
The defining feature in Ekert91 is the suggestion to use the degree of
violation of the CHSH inequality as a test of security.
This conjecture is related to the concept later known as the monogamy of
entanglement \cite{coffman00}: the entanglement between two systems decreases
when a third system (for example, the measurement apparatus of an
eavesdropper) interacts with the pair.

Although BB84 and Ekert91 utilize different aspects of quantum mechanics, once one writes down explicitly the expected qubit states and the measurements that should be performed, the two protocols turn out to generate the same set of correlations \cite{bennett92}. When these calculations were extended to include error correction and privacy amplification, a quantitative link was found between Eve's information (assuming individual attacks) and the amount of violation of the CHSH inequality \cite{fuchs97}, thus vindicating Ekert's intuition. BB84 and Ekert91 came to be considered as fully equivalent. In this perspective, the choice between a prepare-and-measure and an entanglement-based implementation is dictated only by a balance of practical benefits. For instance, BB84 involves an active choice when encoding the logical bits 0 and 1 into the
polarization states, requiring a trusted high-bandwidth random number source
\cite{bienfang04}; in comparison, no active choice is necessary with entanglement-based QKD.
Besides its ability to remove the need for random number generators
\cite{marcikic06}, technical difficulties related to the lack of practical true
single photon sources can be avoided. The price of entanglement-based QKD is a lower key generation rate due to the
limited brightness of contemporary entangled photon-pair sources when compared
with faint coherent pulse approximations of single photon sources.

Recently two theoretical developments pointed to the fact that BB84 and
Ekert91 may not be equivalent after all. The first such development are the
proofs of unconditional security developed by Koashi and Preskill \cite{koa03}
and improved by Ma, Fung and Lo \cite{ma07}. These authors proved that the
security of entanglement-based implementations can be based on the sole
knowledge of the error rate, because this quantity already contains
information about the imperfection of the source --- while such imperfections
(e.g. the photon-number statistics, or spectral distinguishability of
different letters \cite{spectral02}) must be carefully taken into
account in prepare-and-measure schemes \cite{bra00,hwa03,sca04}. The second
development is due to Ac\'{\i}n and coworkers \cite{acin07}. These authors
went back to Ekert's original idea of basing the security \textit{only} on the
combined correlation function $S$ for violating CHSH and derived the formula
\begin{equation}
  I_{Eve}\,=\,h\left({1+\sqrt{S^2/4-1}\over 2}\right)\,,\label{eqacin}
\end{equation}
with the binary entropy $h(x)=-x\log_2(x)-(1-x)\log_2(1-x)$. This formula provides an unconditional security bound under the same assumptions as in \cite{ma07}; it also guarantees partial security in a more paranoid scenario, in which the QKD devices are untrusted (we shall come back to this issue in the conclusions).

In this paper, we describe an entanglement-based QKD experiment in which we
monitor the violation of the CHSH inequality and use (\ref{eqacin}) to quantify
the degree of raw key compression in the privacy amplification
step. Typically, implementations of entanglement-based QKD 
systems do not monitor Bell inequalities \cite{nai00,tit00,marcikic06}; in one
of the first experiments \cite{jennewein00}, a Bell-type inequality was
monitored, but no quantitative measure of security was derived from the
observed violation. 

\begin{figure}
\begin{center}
\includegraphics[width=86mm]{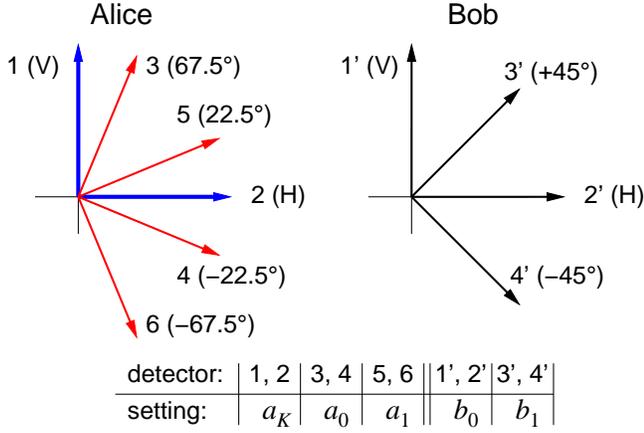}
\end{center}
\caption{Orientation of different detector polarizations. Coincidences (1,1')
  and (2,2') are used for key generation, while coincidences between any of
  (3,4,5,6)  and any of (1',2',3',4') will be used for testing a CHSH
  inequality for various settings.} \label{fig:arrows}
\end{figure}

{\it Experiment.} 
We implement a modified Ekert91 protocol \cite{aci06} that uses a minimal
combination of three detection settings $a_0,a_1,a_K$ on one side, and two distinct
detection settings $b_0,b_1$ on the other side for performing polarization
measurements on a photon-pair in a singlet state $\ket{\Psi^-} = \frac{1}{\sqrt{2}}(\ket{H_A V_B} - \ket{V_A H_B})$.
The setting pair $(a_K,b_0)$  corresponds to
horizontal/vertical polarization, and should lead (in the absence of noise) to
perfectly anti-correlated measurement results which form the raw key. The
setting $b_0$ and the other ones are used to check 
the violation of the CHSH inequality $\left|S\right|\leq 2$ with
\begin{eqnarray}
\label{eq:bell}
S&=& E(a_0,b_0)+E(a_0,b_1)+E(a_1,b_0)-E(a_1,b_1)\,.
\end{eqnarray}
The correlation coefficients $E$ are determined from the number $n_{ij}$ of coincidence events between detectors $i$ on one side and $j$ on the other side, collected during a given integration time $T$. 
Measurement bases are chosen such that a maximal value of $|S|=2\sqrt{2}$ would be expected.  
Basis $b_1$ is chosen to correspond to $\pm45^\circ$ linear polarization, and bases $b,c$ need to form an orthogonal set corresponding to $\pm22.5^\circ,\pm67.5^\circ$ linear polarizations (see Fig.~\ref{fig:arrows}). 
With that, we evaluate for example  
\begin{equation}\label{eq:corrcoeff}
E(a_0,b_0)={n_{3,1'}+n_{4,2'} - n_{3,2'}-n_{4,1'} \over n_{3,1'}+n_{4,2'} +  n_{3,2'}+n_{4,1'} }\,,
\end{equation}
and the other coefficients in (\ref{eq:bell}) accordingly from an ensemble of
pair detection events.

\begin{figure}
\begin{center}
\includegraphics{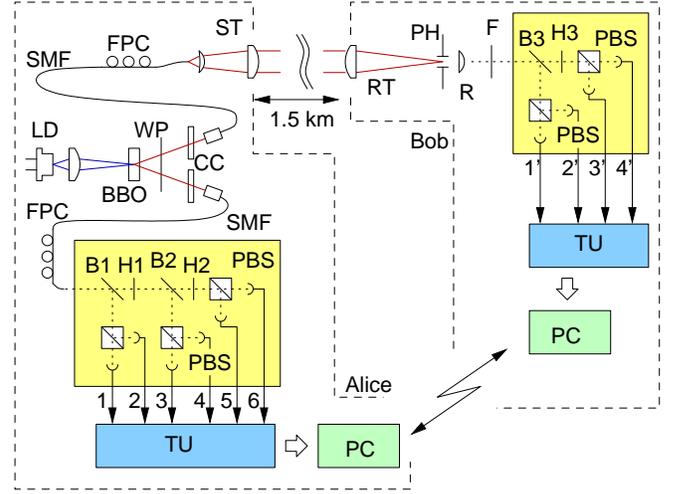}
\end{center}
\caption{Experimental setup. Polarization-entangled photon pairs are generated
  via parametric down conversion pumped by a laser diode (LD) in a
  nonlinear optical crystal (BBO) with walk-off compensation (WP, CC) into
  single mode optical fibers (SMF). A free-space optical channel for one
  detector set (Bob) is realized using small telescopes on both sides (ST,
  RT). Spectral filtering is implemented with a pinhole (PH) which is  imaged
  completely  onto the active area of the photodetectors with a relay lens (R)
  through a  spectral filter (F).  Both parties perform   polarization
  measurements in bases randomly chosen by beam splitters (B1-B3), and defined
  by properly oriented wave plates (H1-H3) in front of polarizing beam
  splitters (PBS) and photon counting detectors. Photo events are registered
  separately with time stamp units (TU) connected to two PC linked via a
  classical channel. } \label{fig:setup}
\end{figure}

The random choice of measurement bases is performed with a combination of polarization-independent beam splitters (B1-B3, see Fig.~\ref{fig:setup}), with a 50:50 splitting ratio. This avoids an explicit generation of a random number by a device. The base settings corresponding to the angles shown in Fig.~\ref{fig:arrows} are adjusted by appropriately oriented half wave plates (H1-H3).

The remaining elements of the experimental setup are similar to a previous experiment implementing an entanglement-based BB84 protocol \cite{marcikic06}.
Polarization-entangled photon pairs are generated in a compact diode-laser
pumped non-collinear type-II parametric down conversion process \cite{kwiat95}
with efficient collection techniques into single mode optical
fibers \cite{kurtsiefer01, trojek04}.
We pump a  2\,mm thick $\beta$-Barium Borate (BBO) crystal at a wavelength of
407\,nm with a power of 40\,mW and observe a photo coincidence rate of about
18000\,s$^{-1}$ in passively-quenched silicon avalanche photodiodes directly
at the source. 
We separate the two measurement devices by $\approx1.5$\,km in an urban
environment, introducing a link 
loss of about 3\,dB caused primarily by atmospheric absorption at the down
converted wavelength of 810\,nm and transmission fluctuations. 

Background light suppression (at night) was accomplished using a spatial
filter (PH) in the receiving telescope (acceptance range
$\Omega=6.5\cdot10^{-9}$\,sr) and a color glass filter (RG780) with a peak
transmission of $\approx90\%$ for the down-converted light at 810\,nm.

Correlated photons are identified by recording their time of arrival at each
detector and running a cross correlation of the timing information on both
sides (similar to the scheme in \cite{marcikic06}). The virtual coincidence
window defined in software was 3.75\,ns, and we monitored the accidental
coincidences in an equally wide time window offset by 20\,ns. Detector time
delay compensation was adjusted to better than 0.5\,ns to avoid leakage
through a classical timing channel \cite{lamas07}.

\begin{figure}
\begin{center}
\includegraphics[width=86mm]{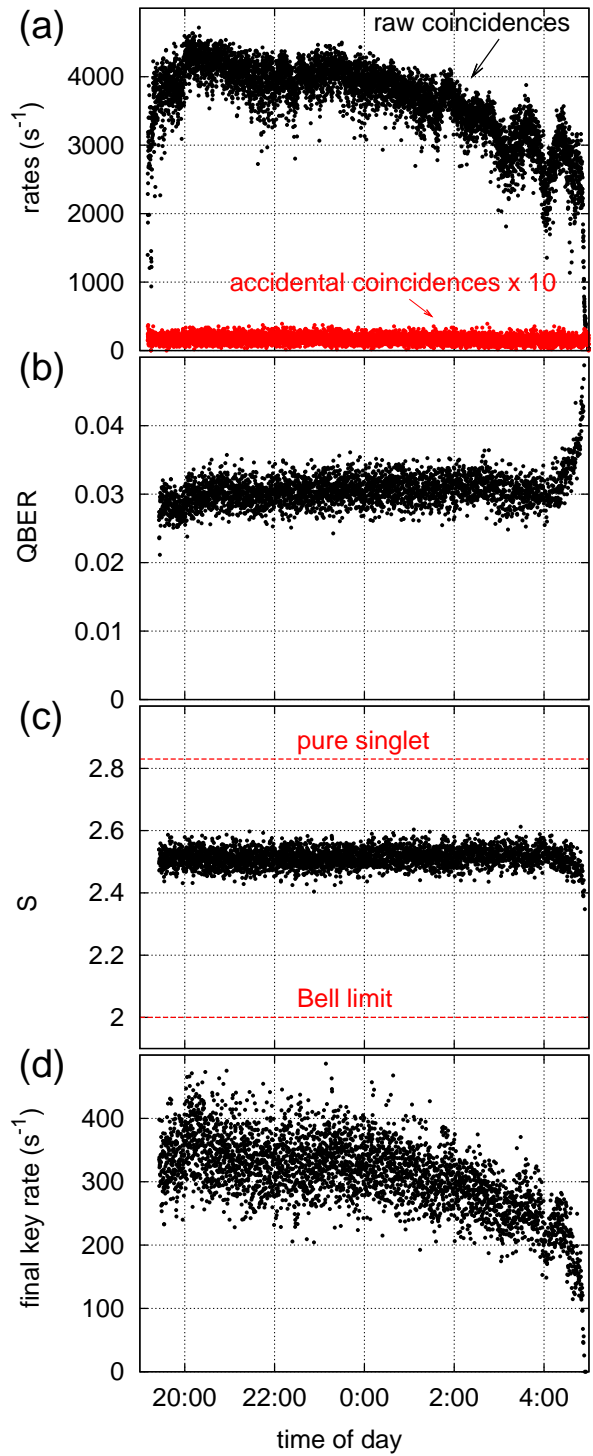}
\end{center}
\caption{Experimental results in a key distribution experiment implementing an
Ekert91 protocol. (a) shows total (upper trace) and accidental (lower trace)
detected coincidence rates between Alice and Bob, (b) the error ration in the
raw key, (c) the degree of violation of a CHSH-type Bell inequality, (d) the
final key rate after error correction and privacy amplification. 
The experiment was terminated by a storm misaligning a telescope of the
optical link at 5\,am.} \label{fig:expresult}
\end{figure}

The experimental results from one 9.5 hour run are shown in
Fig.~\ref{fig:expresult}.  In this interval, we observe a small drop of the
coincidence rate due to an alignment drift in the optical link. Accidental
coincidences were about 0.5\% of the coincidences from down-converted photon
pairs.

Half of the identified photon pairs were seen by detectors
(3,4,5,6) paired with (1',2',3',4'), which were used to evaluate the violation of (\ref{eq:bell}). About a
quarter of the pairs in detector combinations (1,2) with (1',2') contributed
to the raw key, while the residual
quarter of pairs in combinations (1,2) with (3',4') were discarded. Detectors
(1,2,1',2') were adjusted to coincide with the natural
axes of the down conversion crystal to keep the error rate on the raw key as
small as possible. 

Error correction following a modified CASCADE protocol~\cite{cascade} was
performed in real time on packets of at least 10000 raw bits for a targeted
final bit error ratio of $10^{-12}$. The corresponding quantum bit error
ratio (QBER) was extracted out of this procedure
(Fig.~\ref{fig:expresult}b). The combined correlation value $S$ was extracted
via (\ref{eq:corrcoeff}) for that block of raw key, and stayed at around 2.5
over the whole measurement time (Fig.~\ref{fig:expresult}c).  This is not a 
particularly 
high value, and we suspect a broad optical spectrum in the blue pump diode as
a reason for this problem. This is compatible with lower
polarization correlation in the $\pm45^\circ$ basis due to a
residual distinguishability between the two decay paths in the SPDC
process. However, it serves as a typical model for an eavesdropping attempt
e.g. by a partial intercept-resend attack in the H/V basis. While such an
attack is not revealed in the QBER in this protocol, it clearly shows in a
reduction of $S$ from the maximally expected value of $2\sqrt{2}$. 

The average information leakage $l$ per raw bit to an eavesdropper was
estimated for each block following (\ref{eqacin}). Together with the
revealed bits in the error correction procedure (and not assuming that any
errors are due to intrinsic detector noise), we can then establish the secret
key fraction 
Alice and Bob can extract out of the privacy amplification hashing procedure
from a given raw key block. The result over time is shown in
Fig.~\ref{fig:expresult}d, resulting in an average final key rate of around
300\,bit\,s$^{-1}$ or about $10^7$\,bit of error-free secret key.

The estimation of the eavesdropper
knowledge is applicable strictly only for an infinitely large number of bits;
recent work on the security of finite length keys implies that the privacy amplification should be carried out over large ensembles \cite{mawang,hase,scaren1}. For the protocol studied here, a finite-key bound has been presented in Ref.~\cite{scaren2}. By performing privacy amplification on blocks of $n=10^6$ bits of the raw key, the extractable secure key is around half of the asymptotic value \footnote{Note that this value cannot be read from Fig.~1 in \cite{scaren2}, because that plot assumed (i) an a priori relation between $Q$ and $S$ that is not fulfilled in the experiment and (ii) the optimization of the probability of the measurements as a function of $N$, while in the experiment the choices are always made by 50:50 splitters.}.

{\it Conclusion and perspectives.}
We have demonstrated a free space implementation of a modified Ekert91
protocol. The security of the key distilled was derived from the
violation of the CHSH inequality. This ensured that the key was distributed
not by some arbitrary random number generator, but with the non-local
correlations shared by entangled photon-pairs.

Using Ekert91, the authorized parties can give up control over the photon
source. Ac\'{\i}n et al. \cite{acin07} showed that the CHSH violation is in principle sufficient to decide the security (against collective attacks) of a distributed key, even if the measurement apparatus is not trusted. Unfortunately, such a scheme is not yet experimentally feasible because of the stringent requirement
it places on detector efficiencies \cite{acin07,zhao07}. 

A final point must be made about the random choice generator. Our
implementation leaves this choice to the beam splitter B1 in
Fig.~\ref{fig:setup}, which is accessible from the quantum channel. We are then assuming that the eavesdropper cannot change the beam splitter's behavior. This is reasonable; however, it makes our setup fall outside the device-independent scenario, even in the lossless regime. In particular, in that scenario, one can construct a situation in which BB84 would not be secure at all \cite{bb84side} because it is conceivable that different states are sent to the different measurement devices; but this is excluded for a well-behaved beam-splitter, which is precisely our assumption. Device-independent security requires the choice to be made on degrees of freedom independent of those accessible to the eavesdropper.

\section*{Acknowledgment}
This work was supported by the National Research Foundation and the
Ministry of Education, Singapore.

\end{document}